\overfullrule=0pt
\input harvmac
\def\a{{\alpha}}
\def\ad{{\dot \alpha}}

\def\l{{\lambda}}

\def\g{{\gamma}}
\def\d{{\delta}}

\def\s{{\sigma}}

\def\p{{\partial}}

\Title{\vbox{\hbox{ }}}
{\vbox{
\centerline{\bf An Alternative String Theory in Twistor Space}
\centerline{\bf for N=4 Super-Yang-Mills}
}}
\bigskip\centerline{Nathan Berkovits\foot{e-mail: nberkovi@ift.unesp.br}}
\bigskip
\centerline{\it Instituto de F\'\i sica Te\'orica, Universidade Estadual
Paulista}
\centerline{\it Rua Pamplona 145, 01405-900, S\~ao Paulo, SP, Brasil}

\vskip .3in
In this letter, an alternative string theory in twistor space is
proposed for describing perturbative N=4 super-Yang-Mills theory.
Like the recent proposal of Witten, this string theory uses twistor
worldsheet variables and has manifest spacetime superconformal invariance. 
However, in this proposal, tree-level super-Yang-Mills amplitudes come
from open string tree amplitudes as opposed to coming from D-instanton
contributions. 

\vskip .3in

\Date {February 2004}

In a recent paper \ref\witten{
E. Witten, {\it Perturbative Gauge Theory as a String Theory in
Twistor Space}, hep-th/0312171.}, 
Witten has shown that the simple form of maximal 
helicity violating
amplitudes of Yang-Mills theory has a natural generalization
to the non-maximal helicity violating amplitudes. He also constructed 
a topological B-model from twistor worldsheet variables and argued that 
D-instanton contributions in this model reproduce the perturbative
super-Yang-Mills amplitudes. For details on this model and the twistor
approach to super-Yang-Mills, see the review and references in \witten.

The formula for D-instanton contributions of degree $d$ to $n$-gluon
tree-level amplitudes is \witten\ref\googly
{R. Roiban, M. Spradlin and A. Volovich, {\it
A Googly Amplitude from the B-Model in Twistor Space}, hep-th/0402016.}
\eqn\formula{ B(\l_r,\bar\l_r)=
\int d^{2d+2} a ~d^{2d+2} b~ d^{4d+4}\gamma~
\prod_{r=1}^n \int d\sigma_r ~ ( vol(GL(2))^{-1} }
$$\prod_{r=1}^{n-1}
(\s_r -\s_{r+1})^{-1} (\s_n -\s_1)^{-1}
\prod_{r=1}^n \d ({\l_r^2\over\l_r^1} - {{\l^2(\s_r)}\over {\l^1(\s_r)}})
\exp (i\bar\l_r^\ad \l^1_r {{\mu_\ad(\s_r)}\over{\l^1(\s_r)}}) 
$$
$$
Tr [\phi_1({{\psi^A(\s_1)}\over{\l^1(\s_1)}}) \phi_2({{\psi^A(\s_2)}\over{\l^1(\s_2)}}) 
... \phi_n({{\psi^A(\s_n)}\over{\l^1(\s_n)}})]$$
where
$P_r^{\a\ad}=\l_r^\a \bar\l_r^\ad$ is the momentum of the $r^{th}$ state,
$$\l^\a(\s) = \sum_{k=0}^d a_k^\a \s^k,\quad
\mu^\ad(\s) = \sum_{k=0}^d b_k^\ad \s^k,\quad
\psi^A(\s) = \sum_{k=0}^d \g_k^A \s^k,$$
$\phi_r(\psi^A)$ is the N=4 superfield whose lowest component is
the positive helicity gluon and whose top component is
the negative helicity gluon, and the $(vol (GL(2))^{-1}$
factor can be used to remove one of the $a$ integrals and three of
the $\sigma$ integrals.

For maximal helicity violating amplitudes (i.e. $n-2$ positive helicity
gluons and 2 negative helicity gluons), the above formula when $d=1$ has been shown
to give the correct $n$-point amplitude. For non-maximal
helicity violating amplitudes, it has been suggested that this formula
may also give the correct $n$-point amplitude where one has $n-d-1$
positive helicity gluons and $d+1$ negative helicity gluons. Although
there is a possibility that the formula of \formula\ 
needs to be modified for non-maximal amplitudes by contributions from
instantons of lower degree, it has been recently verified 
that no such modifications are necessary when $d=2$
and $n=5$ \googly. It will be assumed below that the formula of \formula\
correctly reproduces the super-Yang-Mills tree amplitudes for any $d$ and $n$.

In this letter, a new string theory in twistor space
is proposed which reproduces
the formula of \formula\ using ordinary open string tree amplitudes as
opposed to D-instanton contributions. This string theory shares many
aspects in common with the orginal idea of Nair in \ref\nair{V.P. Nair,
{\it A Current Algebra for some Gauge Theory Amplitudes}, Phys. Lett.
B214 (1988) 215.}.
The worldsheet matter variables in this
string theory consist
of a left and right-moving set of super-twistor variables,
\eqn\twis{Z_L^I = (\l_L^\a, \mu_L^\ad, \psi_L^A),\quad
Z_R^I = (\l_R^\a, \mu_R^\ad, \psi_R^A)}
for $\a,\ad=1$ to 2 and $A=1$ to 4, a left and right-moving set
of conjugate super-twistor variables,
\eqn\conj{Y_{LI} = (\bar\mu_{L\a}, \bar\l_{L\ad}, 
\bar\psi_{LA}), \quad
Y_{RI} = (\bar\mu_{R\a}, \bar\l_{R\ad}, 
\bar\psi_{RA}),}
and a left and right-moving current algebra,
\eqn\current{j_L^C,\quad j_R^C}
where $C$ is Lie-algebra valued and $j_L^C$ and $j_R^C$
satisfy the usual OPE's of a current algebra, i.e.
\eqn\opej{j_L^C(y) j_L^D(z) \to {{f^{CDE}j_L^E(z)}\over {y-z}} + 
{{k g^{CD}}\over{(y-z)^2}},
\quad
j_R^C(\bar y) j_R^D(\bar z) \to {{f^{CDE}j_R^E(\bar z)}
\over {\bar y-\bar z}}  + 
{{k g^{CD}}\over{(\bar y-\bar z)^2}}.}
The current algebra can be constructed from free fermions, a
Wess-Zumino-Witten model, or any other model.

The worldsheet action is 
\eqn\action{S=\int d^2 z (Y_{LI}\nabla_R Z_L^I + Y_{RI}\nabla_L Z_R^I)
+ S_G}
where $S_G$ is the worldsheet action for the current algebra
and  $(\nabla_R,\nabla_L)$
contains a worldsheet GL(1) connection for which $Z_L^I$ and $Z_R^I$ have charge
$+1$, and $Y_{LI}$ and $Y_{RI}$
have charge $-1$. 

Quantizing this worldsheet action gives rise to left and right-moving
Virasoro ghosts, $(b_L,c_L)$ and $(b_R,c_R)$, as well as left and
right-moving GL(1) ghosts, $(u_L,v_L)$ and $(u_R,v_R)$. The 
untwisted left-moving stress tensor is 
\eqn\stresszero{T_0 = Y_{LI} \p_L Z^I_L + T_G + b_L\p_L c_L + \p_L(b_L c_L) + 
u_L\p_L v_L}
where $T_G$ is the left-moving stress tensor for the current algebra, and the 
left-moving GL(1) current is
\eqn\uone{J = Y_{LI} Z^I_L.}
To have vanishing conformal anomaly, the current algebra must be chosen
such that the central charge contribution from $T_G$ is 28.
Note that there is no GL(1) anomaly because of cancellation between
bosons and fermions in $J$.

The open string theory 
is defined using the conditions
\eqn\boundar{Z_L^I=Z_R^I,\quad Y_{LI}=Y_{RI},
\quad j_L^C=j_R^C,\quad c_L=c_R,\quad b_L=b_R,
\quad v_L=v_R,\quad u_L=u_R} 
on the open string boundary. Unlike a usual open string theory where
Lie algebra indices come from Chan-Paton factors, the Lie algebra indices
in this open string theory come from a current algebra.

The physical integrated and unintegrated
open string vertex operator for the super-Yang-Mills states is
\eqn\vertex{V= \int dz ~j^C(z) \Phi_C(Z(z)), 
\quad U= c(z) j^C(z) \Phi_C(Z(z)).}
The superfields
$\Phi_C(Z)$ are similar to those defined in \witten, namely for
a super-Yang-Mills state with momentum $P_r^{\a\ad}=\l_r^\a\bar\l_r^\ad$,
\eqn\phidef{\Phi_C(Z(z_r))= \d ( {{\l^2_r}\over{\l^1_r}}
- {{\l^2(z_r)}\over{\l^1(z_r)}}) 
\exp (i\bar\l^\ad_r \l^1_r {{\mu^\ad(z_r)}\over
{\l^1(z_r)}}) \phi_C({{\psi^A(z_r)}\over{\l^1(z_r)}})}
where $\phi_C(\psi^A)$ is the same N=4 superfield as in \formula.
Note that $\Phi_C(Z)$ is GL(1)-neutral and has zero conformal weight.

Tree-level open string scattering amplitudes are computed in the usual
manner from the disk correlation function
\eqn\correl{A= \langle U_1(z_1) U_2(z_2) U_3(z_3) \int dz_4 V_4(z_4) ...
\int dz_n V_n (z_n)\rangle}
where different twistings of the stress tensor are used to compute
different helicity violating amplitudes. For amplitudes involving
$(n-d-1)$ positive helicity gluons and $d+1$ negative helicity
gluons, the twisted stress tensor is defined as
\eqn\twisted{T_d = T_0 + {d\over 2} \p J}
where $T_0$ and $J$ are defined in \stresszero\ and \uone.
Note that $T_d$ has no conformal anomaly since $J$ has no GL(1) anomaly.

So after twisting, $Z^I$ has conformal weight $-{d\over 2}$ and
$Y_I$ has conformal weight ${d+2}\over 2$. This means that the disk
correlation function of \correl\ involves an integration over the
$4d+4$ bosonic and $4d+4$ fermionic zero modes of $Z^I$, except
for the one bosonic zero mode which can be removed using the worldsheet GL(1) gauge
invariance.
Performing the correlation function for
the current algebra gives the contribution\foot{
As was pointed out to me 
by Edward Witten, one also gets multitrace contributions
such as
\eqn\subtr{Tr[\phi_1 ...\phi_m]Tr[\phi_{m+1} ...\phi_n]
(\prod_{r=1}^{m-1} (z_r-z_{r+1})^{-1} (z_m-z_1)^{-1})
(\prod_{s=m+1}^{n-1} (z_s-z_{s+1})^{-1} (z_n-z_{m+1})^{-1})}
coming from other contractions of the current algebra. These
multitrace contributions are also present in the amplitudes
coming from D-instantons in \witten\ and in the proposal of Nair in \nair.}
\eqn\contr{Tr[\phi_1 ...\phi_n]\prod_{r=1}^{n-1} (z_r-z_{r+1})^{-1} (z_n-z_1)^{-1},}
and the $(b,c)$ correlation function
gives the factor $(z_1-z_2)(z_2-z_3)(z_3-z_1)$.

So one obtains the formula
\eqn\formulatwo{ A=
\int d^{2d+2} a ~d^{2d+2} b~ d^{4d+4}\gamma ~
\int dz_1 ...\int dz_n  
(Vol(GL(2)))^{-1}}
$$\prod_{r=1}^{n-1}
(z_r -z_{r+1})^{-1} (z_n -z_1)^{-1}
\prod_{r=1}^n \d ({\l_r^2\over\l_r^1} - {{\l^2(z_r)}\over {\l^1(z_r)}})
\exp (i\bar\l_r^\ad \l^1_r {{\mu_\ad(z_r)}\over{\l^1(z_r)}}) $$
$$
Tr [\phi_1({{\psi^A(z_1)}\over{\l^1(z_1)}}) \phi_2({{\psi^A(z_2)}\over{\l^1(z_2)}}) 
... \phi_n({{\psi^A(z_n)}\over{\l^1(z_n)}})]$$
where
$$\l^\a(z) = \sum_{k=0}^d a_k^\a z^k,\quad
\mu^\ad(z) = \sum_{k=0}^d b_k^\ad z^k,\quad
\psi^A(z) = \sum_{k=0}^d \g_k^A z^k,$$
$(a_k^\a, b_k^\ad,\g_k^A)$ are the zero modes of $Z^I$ on a disk, 
and the SL(2) part of GL(2) can be used to fix three of the $z_r$
integrals and reproduce the $(b,c)$ correlation function.
This formula clearly agrees with the formula of \formula\ for
the D-instanton amplitude where the $\s$ variable from the D1-string
worldvolume has been replaced with the $z$ variable from the open string
boundary.

\vskip 15pt
{\bf Acknowledgements:} I would like to thank Cumrun Vafa, Peter Svrcek
and especially
Edward Witten for useful discussions and
the Institute for Advanced Study for their hospitality and financial
support. I would also like to thank
CNPq grant 300256/94-9,
Pronex 66.2002/1998-9, and Fapesp grant 99/12763-0 for partial
financial support. 

\listrefs

\end